\documentstyle[aps,epsfig]{revtex}
\begin{document}
\bibliographystyle{prsty}
\title{Reflection of two-gap nature in penetration depth 
       measurements of MgB$_2$ film}
\author{Mun-Seog Kim, John A. Skinta, and Thomas R. Lemberger}
\address{Department of Physics, Ohio State University, Columbus, OH 43210-1106}

\author{W. N. Kang, Hyeong-Jin Kim, Eun-Mi Choi, and Sung-Ik Lee}
\address{National Creative Research Initiative Center 
         for Superconductivity and Department of Physics, 
         Pohang University of Science and Technology,
         Pohang 790-784, Republic of Korea}
\date{\today}
\maketitle
\begin{abstract}
The magnetic penetration depth, $\lambda(T)$, in the basal plane of
a magnesium diboride (MgB$_2$) film was measured using a two-coil mutual 
inductance technique at 50 kHz. This film has $T_c\simeq 38$ K, 
$\Delta T_c \leq 1$ K, and $\lambda(0)\sim 1500$ \AA.
At low temperatures, $\lambda^{-2}(T)$ shows a clear exponential temperature 
dependence, indicating $s$-wave superconducting order parameter symmetry. 
However, the data are not quantitatively well described by theory assuming 
a single gap. From the data fit by the full BCS calculation assuming a double gap, 
the values of the two distinct gaps were obtained: $\Delta_S(0)=2.61\pm0.41$ meV 
and $\Delta_L(0)=6.50\pm0.33$ meV. The contributions of the small and the large 
gaps to the total superfluid density at $T=0$ were estimated to be 21\% and 79\%, 
respectively. Finally, we consider the effect of gap anisotropy on the penetration 
depth measurements, and find that the gap anisotropy does not play a significant 
role in determining the temperature dependence of the penetration depth.
\end{abstract}

\section{introduction}
Since the discovery of superconductivity in metallic MgB$_2$,\cite{akimitsu_1} 
abundant research has been carried out to elucidate its basic mechanism. 
MgB$_2$ has some notable features which contrast with cuprate superconductors.
First, electron coupling is mediated by phonons, as indicated by the observation
of a prominent isotope effect.\cite{budko_1} However, the transition temperature, 
$T_c\simeq39$K, might be somewhat higher than the theoretical prediction assuming 
a conventional phonon mechanism.\cite{mcmillan_1} Recently, it was suggested 
that anisotropy in electron-phonon couplings plays a significant role in the 
unusually high transition temperature.\cite{liu_1,hjchoi_2}
Secondly, many  theoretical\cite{liu_1,hjchoi_2,hjchoi_1,golubov_1}
and experimental works\cite{schmidt_1,giubileo_1,laube_1,szabo_1,seneor_1,naidyuk_1,%
bouquet_1,y_wang_1,yang_1,walti_1,kotegawa_1,tsuda_1,xk_chen_1}  
suggest that MgB$_2$ has two separate gaps and that the symmetry of each gap 
is $s$-wave with a substantial gap anisotropy. The Fermi surface of MgB$_2$ consists 
of two nearly cylindrical (2D) sheets and two tubular networks (3D).\cite{kortus_1} 
While the value of the gap associated with 3D sheets is in the range 
$0.5\leq\Delta(0)/k_BT_c\leq0.95$, for 2D networks the value is about 
$1.8\leq\Delta(0)/k_BT_c\leq2.2$.
This double-gap structure and its anisotropic nature would be expected 
to play an important role in the physical properties of the compound. 

In the early stages of MgB$_2$ study, a number of groups claimed unconventional 
superconductivity\cite{panagopoulos_1} or $s$-wave order parameter 
symmetry\cite{manzano_1,lamura_1,prozorov_1} from penetration depth measurements 
on various forms of samples. For example, Prozorov {\em et al}.\cite{prozorov_1} showed 
a clear exponential behavior of  $\lambda(T)$ measured using a microwave technique 
for MgB$_2$ wires. On the other hand, Pronin {\em et al}.\cite{pronin_1} reported the 
penetration depth measurements of $c$-axis oriented films with $T_c\simeq32$ K. 
They claimed that the penetration depth shows $T^2$ temperature dependence rather 
than exponential behavior at low temperatures, and suggested a strong gap anisotropy 
or the existence of nodes in the gap as its origin. At present, some of the results 
of  penetration depth measurements and their interpretations are still controversial. 

In this work, we measure the magnetic penetration depth of a high-quality MgB$_2$ film 
via a mutual-inductance technique. At low temperatures, 
the superfluid density, $\rho_s \propto\lambda^{-2}(T)$, shows a 
clear exponential temperature dependence. However, the model assuming a single gap 
does not describe the experimental results. A full calculation within the BCS framework 
assuming existence of a double gap, successfully describes our data. The values of the two gaps 
obtained from fits to the data are consistent  with previous reports. 
Also, the contributions of each gap to the superfluid density are deduced from the 
analysis.

Initially, we compare the data with the low-temperature expansion formula
for $\lambda^{-2}(T)$ assuming one or two gaps to get a rough estimate
of the gap values. Then, we refine the values from a full calculation
of the penetration depth. Finally, we consider how the penetration depth 
is affected by gap anisotropy. 

\section{experimental aspects}
The MgB$_{2}$ thin film was fabricated using a two-step method; the
detailed process is described elsewhere.\cite{wnkang_1} First, an
amorphous boron thin film was deposited on a (1 $\bar{1}$ 0 2) Al$_{2}$O$_{3}$
substrate of $1\times1$ cm$^2$ at room temperature by pulsed laser. 
Then, the boron thin film was put into a Nb tube with high purity 
Mg metal (99.9\%), and the Nb tube was then sealed using an arc furnace in an Ar 
atmosphere. Finally, the heat treatment was carried out at 900$^\circ$C for 10 
to 30 minutes in an evacuated quartz ampoule, which was sealed under high vacuum. 
The film thickness is 0.3 $\mu $m, confirmed by scanning electron 
microscopy. X-ray diffraction patterns indicated that the MgB$_{2}$ thin
film has a highly $c$-axis-oriented crystal structure normal to the substrate
surface; no impurity phase is observed. 

\begin{figure}
\center
\includegraphics[width=10cm,height=7.5cm]{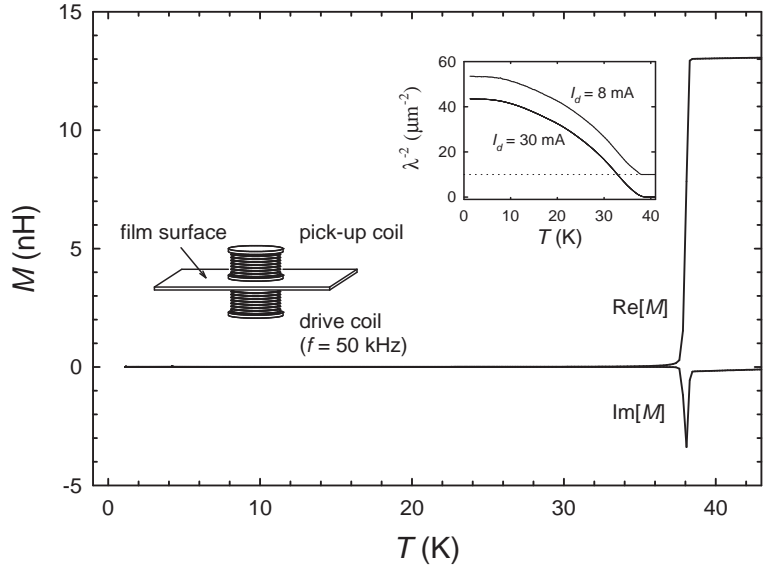}\\
\vspace*{1cm}
\caption{Representative complex mutual inductance $M(T)$ of MgB$_2$ measured using 
two-coil method. Inset (left) shows schematic diagram of measurement configuration.
Inset (right) shows $\lambda^{-2}(T)$ curves extracted from $M(T)$ measured at 
different current levels. Upper curve was shifted by 10 $\mu$m$^{-2}$ for 
comparison.}
\label{fig1}
\end{figure}

The penetration depth, $\lambda(T)$, was measured using a two-coil
mutual inductance technique described in detail elsewhere.\cite{turneaure_1,turneaure_2} 
The MgB$_2$ film is centered between drive and pick-up coils with diameter of $\sim1$ mm.
The inset of Fig.\ \ref{fig1} schematically illustrates the measurement 
configuration. A current (1 mA$\leq I_d\leq$ 30 mA) at 50 kHz in the drive coil induces 
screening currents in the film.
The net magnetic field from the drive coil and the induced current
in film are measured as a voltage across the pick-up coil. 
Because the coils are much smaller than the film, the applied field is
concentrated near the center of the film and demagnetizing effects at the film 
perimeter are not relevant. All data presented here are taken in the linear response 
regime. Figure\ \ref{fig1} shows representative mutual inductance, $M(T)$, data measured 
with $I_d\simeq30$ mA for $T\leq25$ K and $I_d\simeq 2$ mA for $T\geq25$ K.
The mutual inductance technique enables us to extract absolute values 
as well as temperature dependence of the penetration depth from 
the mutual inductance data.

The procedure to extract $\lambda^{-2}(T)$ from $M(T)$ is the following:
First, a constant background (zero position) due to stray couplings between coils 
is subtracted 
from raw data. This constant background can be estimated from measuring at $T=4.2$ K
the mutual inductance of Pb foil with identical shape and area 
as substrate using the same measurement probe. 
In this background measurement,
the magnetic penetration depth of Pb is so small compared to foil 
thickness that no magnetic field goes through film. 
After the subtraction of the background,
the data is normalized to the value of mutual inductance at $T=50$ K  
(initial position). The normalization removes uncertainties associated
with amplifier gains and nonideal aspects of the coil windings.
The subtracted and normalized mutual inductance is converted to 
complex conductivity, $\sigma=\sigma_1-i\sigma_2$, where $\sigma_1$ and 
$\sigma_2$ are real and imaginary parts of the conductivity.
Finally, the penetration depth is determined from the imaginary part of conductivity via
the relationship $\sigma_2=1/\mu_0\omega\lambda^2$, where $\mu_0$ is the 
magnetic permeability of vacuum and $\omega$ is the frequency of drive current.
The accuracy of $\lambda^{-2}$ is limited by 10\% uncertainty in film
thickness. However, the temperature dependence of $\lambda^{-2}$ is unaffected 
by the uncertainty. The other inset of Fig.\ \ref{fig1} displays $\lambda^{-2}(T)$ curves
deduced from $M(T)$ measured at two different $I_d$ levels. 
Although the signal-to-noise ratio of upper curve ($I_d=10$ mA for $T\leq$ 25 K) 
is smaller than that of lower one ($I_d=30$ mA for $T\leq$ 25 K), 
two curves do not show any quantitative difference. 
(Upper curve was shifted by 10 $\mu$m$^{-2}$ for comparison.)

\section{Results and discussion}
\subsection{Theoretical description of data assuming a single gap}
\begin{figure}
\center
\includegraphics[width=10cm,height=7.5cm]{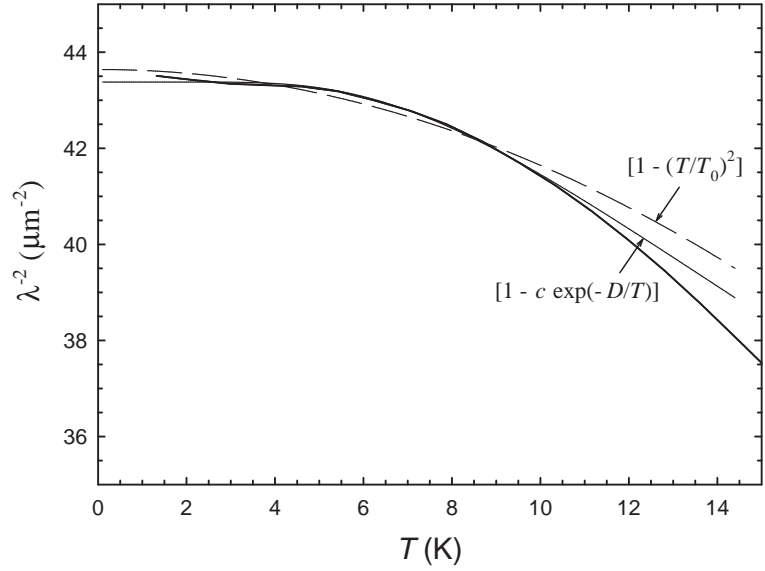}\\
\vspace*{1cm}
\caption{$\lambda^{-2}(T)$ of MgB$_2$ film at low temperatures.
Thin solid and dashed lines denote best fits of $\lambda^{-2}(0)[1-c\exp(-D/T)]$ and 
$\lambda^{-2}(0)[1-(T/T_0)^2]$ to the data, respectively.}
\label{fig2}
\end{figure}

Figure\ \ref{fig2} shows  $\lambda^{-2}(T)$ at temperatures below 15 K. 
The value of $\lambda^{-2}(T)$ at $T\simeq1.3$ K is about 43 $\mu$m$^{-2}$, 
which corresponds to $\lambda\simeq 150$ nm. 
To examine the temperature dependence of $\lambda^{-2}(T)$ at low temperatures, 
we fit first $\sim5$\% drop in $\lambda^{-2}(T)$ to an 
exponential-type function $\lambda^{-2}(T)\sim 1-c \exp(-D/T)$, where $c$ are $D$ are 
adjustable parameters. 
As presented by thin solid line in the figure, the fit is reasonably good.
On the other hand, when we fit the data in the same temperature region to 
a quadratic form $\lambda^{-2}(T)\sim 1-(T/T_0)^2$ as 
in Ref.\ \cite{panagopoulos_1,pronin_1}, the fit deviates 
significantly from the data (dashed line in the same figure).
The $\chi^2$ values for exponential and quadratic fits are $3.96\times10^{-5}$ and 
$5.13\times10^{-4}$, respectively.

\begin{figure}
\center
\includegraphics[width=10cm,height=7.5cm]{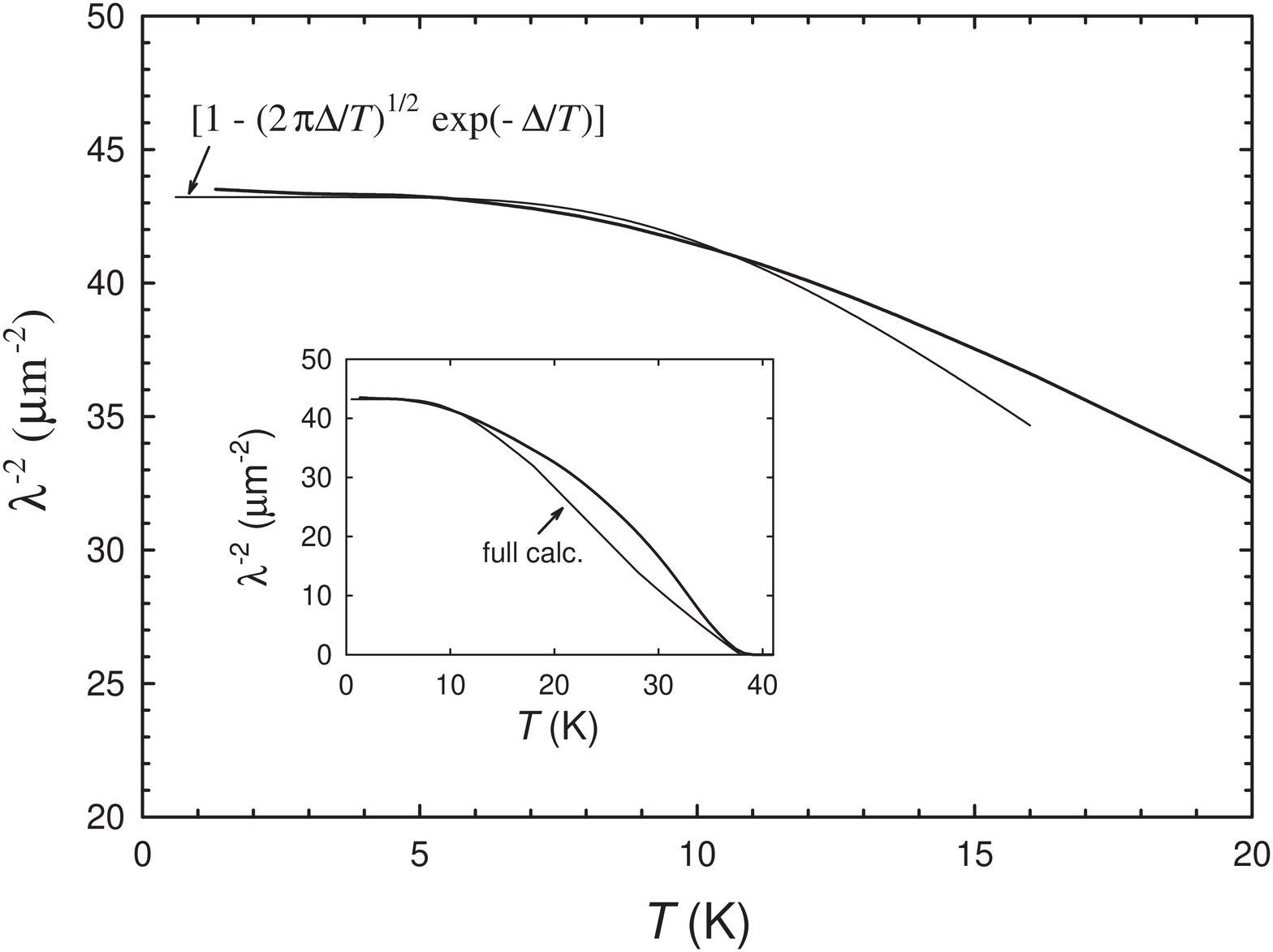}\\
\vspace*{1cm}
\caption{$\lambda^{-2}(T)$ of MgB$_2$ film at low temperatures.
Thin solid line denotes a best fit of Eq.(\ref{eq1}). From this fit, 
$\Delta(0)=4.29$ meV $(\Delta(0)/k_BT_c=1.31)$ was  obtained. Inset: 
$\lambda^{-2}(T)$ for temperatures below $T_c$. Thin solid line is 
a full BCS calculation assuming $\Delta(0)=4.29$ meV.} 
\label{fig3}
\end{figure}

The exponential $T$-dependence of $\lambda^{-2}(T)$ at low temperatures
can be regarded to reflect $s$-wave order-parameter symmetry in this compound.
For clean BCS-type superconductors\cite{tinkham_1}, $\lambda^{-2}(T)$ is given by
	\begin{equation} 
	\frac{\lambda^{-2}(T)}{\lambda^{-2}(0)} = 1 - 2\int_\Delta^\infty 
	\bigg( -\frac{\partial f(E)}{\partial E}\bigg)D(E)dE,
	\label{bcsfull}
	\end{equation}
where $\Delta$ is the superconducting energy gap, $f(E)\equiv[\exp(-E/k_BT)+1]^{-1}$
is the Fermi distribution function,
and $D(E)\equiv E/(E^2-\Delta^2)^{1/2}$, is the quasiparticle density of states (DOS).
This equation can be expanded at low temperatures, where $\Delta$ is nearly constant, 
in the following way\cite{halbritter_1}:
	\begin{equation}
	\frac{\lambda^{-2}(T)}{\lambda^{-2}(0)}
	\simeq 
	1- \bigg(\frac{2\pi\Delta(0)}{k_BT}\bigg)^{1/2} 
	\exp(-\Delta(0)/k_BT),
	\label{eq1}
	\end{equation}
where $\Delta(0)$ is the energy gap at zero temperature. 
Thin solid line in Fig.\ \ref{fig3} represents Eq.(\ref{eq1}) fitted to data. 
The comparison between data and theory is restricted to the first $\sim5$\% drop in 
$\lambda^{-2}(T)$, where the $T$-dependence of $\Delta$ is not significant.
This comparison yields $\Delta(0)=4.29$ meV $[\Delta(0)/k_BT_c=1.31]$ and 
$\lambda^{-2}(0)=43.2$ $\mu$m$^{-2}$.
The gap value is significantly smaller than the BCS weak coupling limit 
$\Delta(0)/k_BT_c \simeq 1.76$. 
Using the gap value deduced above, $\lambda^{-2}(T)$ in the whole 
temperature region below $T_c$ can be obtained by a full BCS calculation 
using Eq.\ (\ref{bcsfull}). 
If the above one-gap fit is valid, the full calculation is expected to describe 
the experimental $\lambda^{-2}(T)$ for the entire temperature region below $T_c$. 
Thin solid line in the inset of Fig.\ \ref{fig3} represents this full calculation. 
The curve does not give a correct description of the data at high 
temperatures. 
While the data show negative curvature at high temperatures, 
the theoretical line shows weakly positive curvature.

\subsection{Theoretical description of data assuming a double gap}
A number of experimental and theoretical groups have proposed the existence of 
two gaps in the DOS of MgB$_2$. The larger gap belongs to the quasi-2D
Fermi surface derived from B-B ($\sigma$) bonds, and the smaller gap belongs to the 
quasi-3D Fermi surface derived from B-Mg-B ($\pi$) bonds. We model this two-gap nature
by writing $D(E)$ as a sum of two BCS DOS. Thus, low-temperature expansion of 
$\lambda^{-2}(T)$ can be expressed by 
	\begin{eqnarray*}
	\lambda^{-2}(T)
	&\simeq&
	\lambda_S^{-2}(0)\Bigg[1-\bigg(\frac{2\pi\Delta_S(0)}{k_BT}\bigg)^{1/2} 
	\exp(-\Delta_S(0)/k_BT)
	\Bigg] \\
	&~&+
	\lambda_L^{-2}(0)\Bigg[1-\bigg(\frac{2\pi\Delta_L(0)}{k_BT}\bigg)^{1/2} 
	\exp(-\Delta_L(0)/k_BT)
	\Bigg],
	\end{eqnarray*}
where the $\Delta_S(0)$ [$\Delta_L(0)$] and $\lambda_S^{-2}(0)$ [$\lambda_L^{-2}(0)$]
are the value of the small (large) gap and the contribution of small (large) gap
to total superfluid density ($\propto\lambda^{-2}$), respectively. 
For comparison with data, it is more convenient to convert the above equation to 
the following form
	\begin{eqnarray}
	\frac{\lambda^{-2}(T)}{\lambda^{-2}(0)}
	&\simeq& 
	1-c_1\bigg(\frac{2\pi\Delta_S(0)}{k_BT}\bigg)^{1/2} 
                          \exp(-\Delta_S(0)/k_BT)  \nonumber \\
	&~&-c_2\bigg(\frac{2\pi\Delta_L(0)}{k_BT}\bigg)^{1/2}  
                          \exp(-\Delta_L(0)/k_BT),
	\label{eq2}
	\end{eqnarray}
where $\lambda^{-2}(0)=\lambda_{S}^{-2}(0)+\lambda_{L}^{-2}(0)$, 
$c_1=\lambda_{S}^{-2}(0)/\lambda^{-2}(0)$, and $c_2=(1-c_1)$.

\begin{figure}
\center
\includegraphics[width=10cm,height=7.5cm]{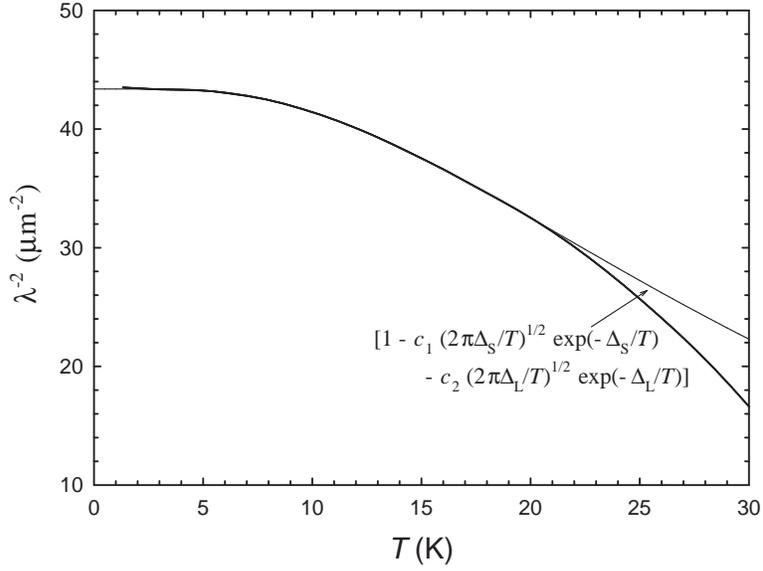}\\
\vspace*{1cm}
\caption{$\lambda^{-2}(T)$ fitted 
by Eq.(\ref{eq2}). In this fit, the values of two distinct gaps 
$\Delta_S(0)=2.57$ meV and $\Delta_L(0)=5.82$ meV were obtained.} 
\label{fig4}
\end{figure}

At very low temperatures, the change of superfluid density 
with  temperature, i.e., the quasiparticle excitation, is dominated by
the small gap. In other words, the role of the large gap in $\lambda^{-2}(T)$
is relevant at higher temperatures. Thus, we extend the fitting region 
up to $\sim15$\% drop in $\lambda^{-2}(T)$. 15\% drop in $\lambda^{-2}(T)$ 
corresponds to about 10\% drop in $\Delta(T)/\Delta(0)$ in the BCS weak 
coupling limit. Accordingly, in this fit about 10\% error due to change of 
$\Delta$ can be expected. 
Figure\ \ref{fig4} shows the comparison of Eq.\ (\ref{eq2}) with data. 
From this fit, we obtains two distinct gap values, $\Delta_S(0)=2.57$ meV and 
$\Delta_L(0)=5.82$ meV, corresponding to $\Delta_S(0)/k_BT_c=0.79$ and 
$\Delta_L(0)/k_BT_c=1.78$. In the case of the small gap, 
the value is consistent with previous reports. But the large gap value is somewhat 
smaller than those in the literatures (Table\ \ref{table1}). 

The success of the fit motivates a full calculation in the extended temperature range
for a more precise description of the data. We assume isotropic $s$-wave gaps
on the two pieces of Fermi surface, and perform a full calculation of $\lambda^{-2}(T)$ 
according to

	\begin{equation}
	\frac{\lambda^{-2}(T)}{\lambda^{-2}(0)} =  
	1 - 2\bigg[
 	c_1\int_{\Delta_S}^\infty \bigg(-\frac{\partial f}{\partial E}\bigg)D_S(E)dE 
	+c_2\int_{\Delta_L}^\infty \bigg(-\frac{\partial f}{\partial E}\bigg)D_L(E)dE
	\bigg],
	\label{eq3}
	\end{equation}
where $c_1$ is adjustable parameter which determines the contribution of the small 
gap to the superfluid density and $c_2=(1-c_1)$. 

\begin{figure}
\center
\includegraphics[width=10cm,height=7.5cm]{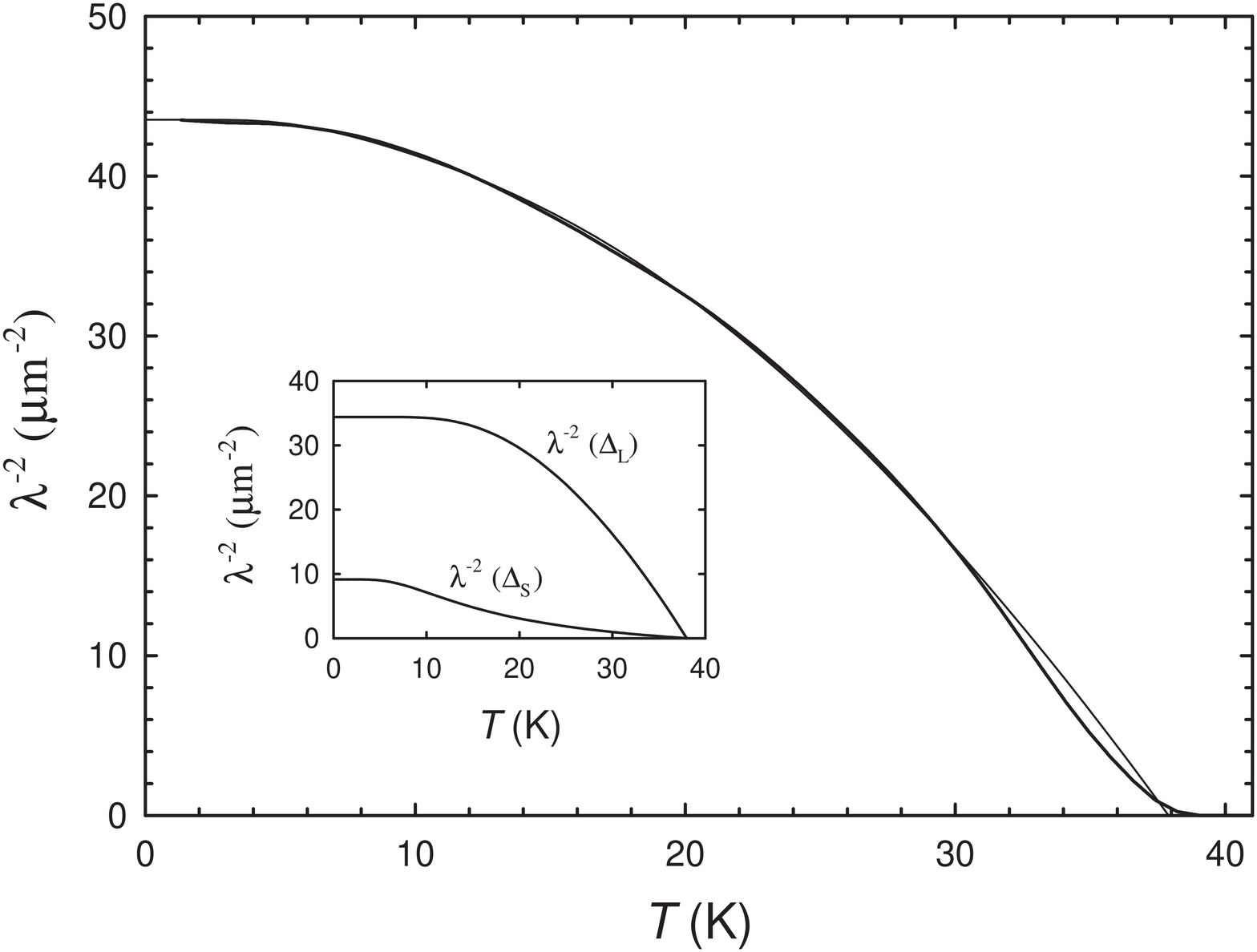}\\
\vspace*{1cm}
\caption{$\lambda^{-2}(T)$ for whole temperature range below $T_c$. Thin solid line is 
a full BCS calculation of $\lambda^{-2}(T)$ assuming two distinct gaps. 
From this, refined gaps values $\Delta_S(0)=2.61$ meV and $\Delta_L(0)=6.50$ meV 
were obtained. Inset: Theoretical curves of $\lambda^{-2}(T)$. Upper and lower curves
are the contributions of the large ($\Delta_L(0)=6.50$ meV) and 
the small ($\Delta_S(0)=2.61$ meV) gaps, respectively.}
\label{fig5}
\end{figure}

Figure\ \ref{fig5} shows our attempt to fit the data using Eq.(\ref{eq3}). Except 
near $T_c$, the theoretical line gives a good fit to the data. From this, we obtain 
the gap values $\Delta_S(0)=2.61\pm0.41$ meV and $\Delta_L(0)=6.50\pm0.33$ meV. These  
are fairly consistent with previous reports (Table I). Also, the contributions of 
each gap to $\lambda^{-2}(0)$, i.e., $c_2=0.79\pm0.06$ is deduced.
The inset of Fig.\ \ref{fig5} shows theoretical $\lambda^{-2}(T)$ curves,
where the contributions of each gap are separately plotted.

In the above analysis, we described the $\lambda^{-2}(T)$ theoretically assuming 
the gaps ($\Delta_L$ and $\Delta_S$) being isotropic on the Fermi surfaces.
According to a recent theoretical calculation, the values of the small and large gaps 
are distributed in the range of 1 meV $\leq\Delta_S \leq 3$ meV 
and 6.5 meV $\leq \Delta_L \leq 7.5$ on the Fermi surfaces.\cite{hjchoi_1} 
Here we suppose two phenomenological models for gap distribution. In the first model,
the gap is distributed uniformly around the average value of gap, $\Delta_0$.
In the alternative model, we assume the normal (Gaussian) distribution of the gap.
Using these models, we calculate theoretical curves of $\lambda^{-2}(T)$ assuming 
$\sim\pm25$\% variation of gap around $\Delta_0$ on the Fermi surface. 
The gap distribution of $\pm25$\% is sufficient to account for real gap anisotropy 
in MgB$_2$.\cite{naidyuk_1} The calculations reveal that the change in $\lambda^{-2}(T)$ due 
to the gap distribution is not significant. In fact, the maximum change in 
$\lambda^{-2}$ due to the gap anisotropy is only about 2\% in the case of 
the uniform distribution. The normal gap distribution causes 
negligibly small change in $\lambda^{-2}$. These results lead us to the 
conclusion that the gap anisotropy on the Fermi surface  of MgB$_2$ 
is not relevant in determining the temperature dependence of the penetration depth. 

\section{Summary}
The magnetic penetration depth $\lambda(T)$ of a high-quality, $c$-axis oriented 
MgB$_2$ film was obtained from mutual-inductance measurements in the linear-response 
regime. The exponential temperature dependence of $\lambda^{-2}(T)$ at low temperatures 
suggests a nodeless gap on the Fermi surface. However, the data could not be described 
by the $s$-wave theory assuming a single gap even at low temperatures. On the other 
hand, the data were successfully described by the full calculation of $\lambda^{-2}(T)$ 
with two distinct gap values: $\Delta_S(0)=2.61\pm0.41$ meV and 
$\Delta_L(0)=6.50\pm0.33$ meV. At $T=0$, the contribution of the small gap to the 
superfluid density  was found to be 21\%. Finally, two phenomenological models to 
account for gap-size distribution on the Fermi surface were considered. It was found 
that gap-size distribution in MgB$_2$ does not play a significant role in determining 
the temperature dependence of the penetration depth.

\acknowledgments
This work was supported in part by DOE Grant DE-FG02-90ER45427
through the Midwest Superconductivity Consortium (MSK, JAS, and, TRL) and by
Creative Research Initiatives of the Korean Ministry of 
Science and Technology (WNK, HJK, EMC, and SIL).

\begin{table}[h]
\caption{Summary of previously reported superconducting gap values 
         of MgB$_2$ superconductor.}
\begin{tabular}{ccll}
$\Delta_S$ ($\Delta$), meV & $\Delta_L$, meV  & Tool &  Refs. \\ \hline
$2.61\pm0.41$&   $6.5\pm0.33$ & Penetration depth    &  this work\\ 
$1\sim3$     &   $6.5\sim7.5$ & First principle calc.& Ref.[5]  \\
2.5          &   -            & Tunneling      & Ref.[8]  \\ 
3.8          &   7.8          & STM      & Ref.[9]  \\
1.7          &   7            & Point-contact spec.  & Ref.[10] \\
2.8          &   7            & Point-contact spec.  & Ref.[11] \\ 
$2.45\pm0.15$&   $7.0\pm0.45$ & Point-contact spec.  & Ref.[13] \\ 
1.7          &   5.6          & Photoemission spec.  & Ref.[19] \\
2.7          &   6.2          & Raman spec.      & Ref.[20] \\ 
$2.8\pm0.4$  &   -            & Penetration depth    & Ref.[23] \\ 
2.61         &   -            & Penetration depth    &  Ref.[25] 
\end{tabular}
\label{table1}
\end{table}


\begin{references}

\bibitem{akimitsu_1}
J.~Nagamatsu, N.~Nakagawa, T.~Muranaka, Y.~Zenitani, and J.~Akimitsu, Nature
  {\bf 410},  63  (2001).

\bibitem{budko_1}
S.~L. Bud'ko, G.~Lapertot, C.~Petrovic, C.~E. Cunningham, N.~Anderson, and
  P.~C. Canfield, Phys. Rev. Lett. {\bf 86},  1877  (2001).

\bibitem{mcmillan_1}
W.~L. McMillan and J.~M. Rowell,  in {\em Superconductivity}, edited by R.~D.
  Parks (Marcel Dekker, New York, 1969), p.\ 561.

\bibitem{liu_1}
A.~Y. Liu, I.~I. Mazin, and J.~Kortus, Phys. Rev. Lett. {\bf 87},  087005
  (2001).

\bibitem{hjchoi_2}
H.~J. Choi, D.~Roundy, H.~Sun, M.~L. Cohen, and S.~G. Louie, cond-mat/0111183,
  2001.

\bibitem{hjchoi_1}
H.~J. Choi, D.~Roundy, H.~Sun, M.~L. Cohen, and S.~G. Louie, cond-mat/0111182,
  2001.

\bibitem{golubov_1}
A.~A. Golubov, J.~Kortus, O.~V. Dolgov, O.~Jepsen, Y.~Kong, O.~K. Andersen,
  B.~J. Gibson, K.~Ahn, and R.~K. Kremer, cond-mat/0111262, 2001.

\bibitem{schmidt_1}
H.~Schmidt, J.~F. Zasadzinski, K.~E. Gray, and D.~G. Hinks, Phys. Rev. B {\bf
  63},  220504  (2001).

\bibitem{giubileo_1}
F.~Giubileo, D.~Roditchev, W.~Sacks, R.~Lamy, D.~X. Thanh, J.~Klein,
  S.~Miraglia, D.~Fruchart, J.~Marcus, and Ph. Monod, Phys. Rev. Lett. {\bf
  87},  177008  (2001).

\bibitem{laube_1}
F.~Laube, G.~Goll, J.~Hagel, H.~v.~Lohneysen, D.~Ernst, and T.~Wolf,
  cond-mat/0106407, 2001.

\bibitem{szabo_1}
P.~Szab\'{o}, P.~Samuely, J.~Ka\v{c}mar\v{c}ik, T.~Klein, J.~Marcus,
  D.~Fruchart, S.~Miraglia, C.~Marcenat, and A.~G.~M. Jansen, Phys. Rev. Lett.
  {\bf 87},  137005  (2001).

\bibitem{seneor_1}
P.~Seneor, C.-T. Chen, N.-C. Yeh, R.~P. Vasquez, L.~D. Bell, C.~U. Jung,
  Min-Seok Park, Heon-Jung Kim, W.~N. Kang, and Sung-Ik Lee, Phys. Rev. B {\bf
  65},  012505  (2001).

\bibitem{naidyuk_1}
Yu.~G. Naidyuk, I.~K. Yanson, L.~V. Tyutrina, N.~L. Bobrov, P.~N. Chubov, W.~N.
  Kang, H.-J. Kim, E.-M. Choi, and S.-I. Lee, cond-mat/0112452, 2002.

\bibitem{bouquet_1}
F.~Bouquet, R.~A. Fisher, N.~E. Phillips, D.~G. Hinks, and J.~D. Jorgensen,
  Phys. Rev. Lett. {\bf 87},  047001  (2001).

\bibitem{y_wang_1}
Y.~Wang, T.~Plackowski, and A.~Junod, Physica C {\bf 355},  179  (2001).

\bibitem{yang_1}
H.~D. Yang, J.-Y. Lin, H.~H. Li, F.~H. Hsu, C.~J. Liu, S.-C. Li, R.-C. Yu, and
  C.-Q. Jin, Phys. Rev. Lett. {\bf 87},  167003  (2001).

\bibitem{walti_1}
Ch. W\"{a}lti, E.~Felder, C.~Degen, G.~Wigger, R.~Monnier, B.~Delley, and H.~R.
  Ott, Phys. Rev. B {\bf 64},  172515  (2001).

\bibitem{kotegawa_1}
H.~Kotegawa, K.~Ishida, Y.~Kitaoka, T.~Muranaka, and J.~Akimitsu, Phys. Rev.
  Lett. {\bf 87},  127001  (2001).

\bibitem{tsuda_1}
S.~Tsuda, T.~Yokoya, T.~Kiss, Y.~Takano, K.~Togano, H.~Kito, H.~Ihara, and
  S.~Shin, Phys. Rev. Lett. {\bf 87},  177006  (2001).

\bibitem{xk_chen_1}
X.~K. Chen, M.~J. Konstantinovic, J.~C. Irwin, D.~D. Lawrie, and J.~P. Franck,
  Phys. Rev. Lett. {\bf 87},  157002  (2001).

\bibitem{kortus_1}
J.~Kortus, I.~I. Mazin, K.~D. Belashchenko, V.~P. Antropov, and L.~L. Boyer,
  Phys. Rev. Lett. {\bf 86},  4656  (2001).

\bibitem{panagopoulos_1}
C.~Panagopoulos, B.~D. Rainford, T.~Xiang, C.~A. Scott, M.~Kambara, and I.~H.
  Inoue, Phys. Rev. B {\bf 64},  094514  (2001).

\bibitem{manzano_1}
F.~Manzano, A.~Carrington, N.~E. Hussey, S.~Lee, A.~Yamamoto, and S.~Tajima,
  Phys. Rev. Lett. {\bf 88},  047002  (2002).

\bibitem{lamura_1}
G.~Lamura, E.~Di Gennaro, M.~Salluzzo, A.~Andreone, J.~Le Cochec, A.~Gauzzi,
  C.~Cantoni, M.~Paranthaman, D.~K. Christen, H.~M. Christen, G.~Giunchi, and
  S.~Ceresara, Phys. Rev. B {\bf 65},  020506  (2002).

\bibitem{prozorov_1}
R.~Prozorov, R.~W. Giannetta, S.~L. Bud'ko, and P.~C. Canfield, Phys. Rev. B
  {\bf 64},  180501  (2001).

\bibitem{pronin_1}
A.~V. Pronin, A.~Pimenov, A.~Loidl, and S.~I. Krasnosvobodtsev, Phys. Rev.
  Lett. {\bf 87},  097003  (2001).

\bibitem{wnkang_1}
W.~N. Kang, H.-J. Kim, E.-M. Choi, C.~U. Jung, and S.-I. Lee, Science {\bf
  292},  1521  (2001).

\bibitem{turneaure_1}
S.~J. Turneaure, E.~R. Ulm, and T.~R. Lemberger, J. Appl. Phys. {\bf 79},  4221
   (1996).

\bibitem{turneaure_2}
S.~J. Turneaure, A.~A. Pesetski, and T.~R. Lemberger, J. Appl. Phys. {\bf 83},
  4334  (1998).

\bibitem{tinkham_1}
M.~Tinkham, {\em Introduction to Superconductivity}, 2nd ed. (McGraw-Hill, New
  York, 1996).

\bibitem{halbritter_1}
J.~Halbritter, Z. Physik {\bf 243},  201  (1971).

\end{references}

\end{document}